\documentclass[11pt]{article}
\usepackage{exscale}\usepackage{graphicx}
\def\d{\mbox{\rm d}}
\begin{document}
\title{Lagrangians for
 dissipative nonlinear oscillators:
the method of Jacobi Last Multiplier}
 \author{M.C. Nucci  $\;$ and $\;$ K.M. Tamizhmani\footnote{Permanent address: Department of
Mathematics, Pondicherry University, R. V. Nagar, Kalapet,
Pondicherry, 605 014,  INDIA, e-mail: tamizh@yahoo.com}}
\date{Dipartimento di Matematica
e Informatica,Universit\`a di Perugia, 06123 Perugia, Italy,
e-mail: nucci@unipg.it}

 \maketitle
 \begin{abstract}
We present a method devised by Jacobi to derive Lagrangians of any
second-order differential equation: it consists in finding a
Jacobi Last Multiplier. We illustrate the easiness and the power
of Jacobi's method by applying it to several equations and also a
class of equations studied by Musielak with his own method
[Musielak ZE, Standard and non-standard Lagrangians for
dissipative dynamical systems with variable coefficients. J. Phys.
A: Math. Theor. 41 (2008) 055205 (17pp)], and in particular to a
Li\`enard type nonlinear oscillator, and a second-order Riccati
equation.
 \end{abstract}
\section{Introduction}
It should be well-known that the knowledge of a Jacobi Last
Multiplier always yields a Lagrangian of any second-order ordinary
differential equation \cite{JacobiVD}, \cite{Whittaker}. Yet many
distinguished scientists seem to be unaware of this classical
result. In this paper we present again the method of the Jacobi
Last Multiplier in order to compare the easiness and the power of
Jacobi's method with that proposed by Museliak et al
\cite{musetal08} for the same purpose. We have already presented
the properties of the Jacobi Last Multiplier  in \cite{jlm05}.
Some of  our papers \cite{ennity}-\cite{nuctam_1lag} and the
references within may give an idea of the many fields of
applications yielded by Jacobi Last Multiplier.

In \cite{musetal08} the authors searched for a Lagrangian of the
following second-order ordinary differential equation
\begin {equation}
\ddot x+b(x)\dot x^2+c(x)x=0 \label{eq1}
\end{equation}
with $b(x),c(x)$  arbitrary functions of the dependent variable
$x=x(t)$. After some lengthy calculations they found one
Lagrangian. In \cite{nuctam_1lag}  we show that (\ref{eq1}) is a
subcase of a more general class of equations studied by Jacobi
\cite{Jacobi 45}, i.e.
\begin{equation}
\ddot x+\frac{1}{2}\frac{\partial \varphi}{\partial x}\dot
x^2+\frac{\partial \varphi}{\partial t}\dot x+B=0\label{Jeq}
\end{equation}
with $\varphi, B$ arbitrary functions of $t$ and $x$. We applied
Jacobi's method to equation (\ref{eq1}) and showed that many (an
infinite number of) Lagrangians can be easily derived.

In the present paper we show how to obtain many (an infinite
number of) Lagrangians for the class of equations
\begin {equation}
\ddot x+f(x)\dot x+g(x)=0 \label{2}
\end{equation}
with $f(x)$ and $g(x)$ arbitrary functions of the dependent
variable $x(t)$.   With the help of the Jacobi Last Multiplier
standard and nonstandard Lagrangians can be derived without much
effort. In \cite{musetal08_2} the author applied  his lengthy
method to (\ref{2}) in order to obtain at least one Lagrangian.

This paper is organized in the following way. In  section 2, we
illustrate the Jacobi Last Multiplier and its properties
\cite{Jacobi 42}-\cite{JacobiVD}, its connection to Lie symmetries
\cite{Lie1874}, \cite{Lie 12}, and its link to the Lagrangian of
any second-order differential equations \cite{JacobiVD},
\cite{Whittaker}. We also exemplify Jacobi's method with an
equation \cite{MatLak74} of the class of equations (\ref{eq1}),
i.e.:
\begin {equation}
\ddot x=x\frac{ - a + \lambda \dot x^2}{\lambda x^2 + 1}
\label{careq}
\end{equation}
and a nonautonomous equation \cite{NorMar} of the more general
class of equations (\ref{Jeq}), i.e.:
\begin {equation}
\ddot x=-\frac{\dot x^2}{x}+\frac{\dot x}{t}\label{NMeq}.
\end{equation}
Both examples  were  not included in \cite{nuctam_1lag}. In
section 3, we apply Jacobi's method to the class of equations
(\ref{2}), and show some particular examples such as a Li\`enard
type nonlinear oscillator, and a second-order Riccati equation. In
section 4, we conclude with some final remarks.

 In this paper we employ ad hoc interactive programs
\cite{man2} written
in REDUCE language to calculate the Lie symmetry algebra of the equations we study.\\

\section{The method by Jacobi}
The method of the Jacobi Last Multiplier \cite{Jacobi 44
a}-\cite{JacobiVD}) provides a means to determine all the
solutions of the partial differential equation
\begin {equation}
\mathcal{A}f = \sum_{i = 1} ^n a_i(x_1,\dots,x_n)\frac {\partial
f} {\partial x_i} = 0 \label {2.1}
\end {equation}
or its equivalent associated Lagrange's system
\begin {equation}
\frac {\d x_1} {a_1} = \frac {\d x_2} {a_2} = \ldots = \frac {\d
x_n} {a_n}.\label {2.2}
\end {equation}
In fact, if one knows the Jacobi Last Multiplier and all but one
of the solutions, then the last solution can be obtained by a
quadrature. The Jacobi Last Multiplier $M$ is given by
\begin {equation}
\frac {\partial (f,\omega_1,\omega_2,\ldots,\omega_{n- 1})}
{\partial (x_1,x_2,\ldots,x_n)}
 = M\mathcal{A}f, \label {2.3}
\end {equation}
where
\begin {equation}
\frac {\partial (f,\omega_1,\omega_2,\ldots,\omega_{n- 1})}
{\partial (x_1,x_2,\ldots,x_n)} = \mbox {\rm det}\left [
\begin {array} {ccc}
\displaystyle {\frac {\partial f} {\partial x_1}} &\cdots &\displaystyle {\frac {\partial f} {\partial x_n}}\\
\displaystyle {\frac {\partial\omega_1} {\partial x_1}} & &\displaystyle {\frac {\partial\omega_1} {\partial x_n}}\\
\vdots & &\vdots\\
\displaystyle {\frac {\partial\omega_{n- 1}} {\partial x_1}}
&\cdots &\displaystyle {\frac {\partial\omega_{n- 1}} {\partial
x_n}}
\end {array}\right] = 0 \label {2.4}
\end {equation}
and $\omega_1,\ldots,\omega_{n- 1} $ are $n- 1 $  solutions of
(\ref {2.1}) or, equivalently, first integrals of (\ref {2.2})
independent of each other. This means that  $M$ is a function of
the variables $(x_1,\ldots,x_n)$ and  depends on the chosen $n-1$
solutions, in the sense that it varies as they vary. The essential
properties of the Jacobi Last Multiplier are:
\begin{description}
\item{ (a)} If one selects a different set of $n-1$ independent
solutions $\eta_1,\ldots,\eta_{n-1}$ of equation (\ref {2.1}),
then the corresponding last multiplier $N$ is linked to $M$ by the
relationship:
$$
N=M\frac{\partial(\eta_1,\ldots,\eta_{n-1})}{\partial(\omega_1,
\ldots,\omega_{n-1})}.
$$
\item{ (b)} Given a non-singular transformation of variables
$$
\tau:\quad(x_1,x_2,\ldots,x_n)\longrightarrow(x'_1,x'_2,\ldots,x'_n),
$$
\noindent then the last multiplier $M'$ of  $\mathcal{A'}F=0$ is
given by:
$$
M'=M\frac{\partial(x_1,x_2,\ldots,x_n)}{\partial(x'_1,x'_2,\ldots,x'_n)},
$$
where $M$ obviously comes from the $n-1$ solutions of
$\mathcal{A}F=0$ which correspond to those chosen for
$\mathcal{A'}F=0$ through the inverse transformation $\tau^{-1}$.
\item{ (c) } One can prove that each multiplier $M$ is a solution
of the following
 linear partial differential equation: \begin {equation}
\sum_{i = 1} ^n \frac {\partial (Ma_i)} {\partial x_i} = 0; \label
{2.5} \end {equation} \noindent viceversa every solution $M$ of
this equation is a Jacobi Last Multiplier.
\item{ (d) } If one
knows two Jacobi Last Multipliers $M_1$ and $M_2$ of equation
(\ref {2.1}), then their ratio is a solution $\omega$ of (\ref
{2.1}), or, equivalently,  a first integral of (\ref {2.2}).
Naturally the ratio may be quite trivial, namely a constant.
Viceversa the product of a multiplier $M_1$ times any solution
$\omega$ yields another last multiplier
$M_2=M_1\omega$.\end{description}
  Since the existence of a solution/first integral
is consequent upon the existence of symmetry, an alternate
formulation in terms of symmetries was provided by Lie \cite {Lie
12}. A clear treatment of the formulation in terms of
solutions/first integrals  and symmetries
 is given by Bianchi \cite {Bianchi 18}. If we know
$n- 1 $ symmetries of (\ref {2.1})/(\ref {2.2}), say
\begin {equation}
\Gamma_i =
\sum_{j=1}^{n}\xi_{ij}(x_1,\dots,x_n)\partial_{x_j},\quad i = 1,n-
1, \label {2.6}
\end {equation}
Jacobi's last multiplier is given by $M =\Delta ^ {- 1} $,
provided that $\Delta\not = 0 $, where
\begin {equation}
\Delta = \mbox {\rm det}\left [
\begin {array} {ccc}
a_1 &\cdots & a_n\\
\xi_{1,1} & &\xi_{1,n}\\
\vdots & &\vdots\\
\xi_{n- 1,1}&\cdots &\xi_{n- 1,n}
\end {array}\right]. \label {2.8}
\end {equation}
There is an obvious corollary to the results of Jacobi mentioned
above. In the case that there exists a constant multiplier, the
determinant is a first integral.  This result is potentially very
useful in the search for first integrals of systems of ordinary
differential equations.  In particular, if each component of the
vector field of the equation of motion is missing the variable
associated with that component, i.e., $\partial a_i/\partial x_i =
0 $, the last multiplier is a constant, and any other Jacobi Last
Multiplier is a first integral.

Another property of the Jacobi Last Multiplier is  its (almost
forgotten) relationship with the Lagrangian, $L=L(t,x,\dot x)$,
for any second-order equation
\begin{equation}
\ddot x=F(t,x,\dot x) \label{geno2}
\end{equation}
is \cite{JacobiVD}, \cite{Whittaker}
\begin{equation}
M=\frac{\partial^2 L}{\partial \dot x^2} \label{relMLo2}
\end{equation}
where $M=M(t,x,\dot x)$ satisfies the following equation
\begin{equation} \frac{{\rm d}}{{\rm d} t}(\log M)+\frac{\partial F}{\partial
\dot x} =0.\label{Meq}
\end{equation}
Then equation (\ref{geno2}) becomes the Euler-Lagrangian equation:
\begin{equation}
-\frac{{\rm d}}{{\rm d} t}\left(\frac{\partial L}{\partial \dot
x}\right)+\frac{\partial L}{\partial x}=0. \label{ELo2}
\end{equation}
The proof is given by taking the derivative of (\ref{ELo2}) by
$\dot x$ and showing that this yields (\ref{Meq}).
 If one knows a Jacobi Last Multiplier, then $L$ can be
easily obtained by a double integration, i.e.:
\begin{equation}
L=\int\left (\int M\, {\rm d} \dot x\right)\, {\rm d} \dot
x+f_1(t,x)\dot x+f_2(t,x), \label{lagrint}
\end{equation}
where $f_1$ and $f_2$ are functions of $t$ and $x$ which have to
satisfy a single partial differential equation related to
(\ref{geno2}) \cite{laggal}. As it was shown in \cite{laggal},
$f_1, f_2$ are related to the gauge function $G=G(t,x)$. In fact,
we may assume
\begin{eqnarray}
f_1&=&  \frac{\partial G}{\partial x}\nonumber\\
f_2&=& \frac{\partial G}{\partial t} +f_3(t,x) \label{gf1f2o2}
\end{eqnarray}
where $f_3$ has to satisfy the mentioned partial differential
equation and $G$ is obviously arbitrary. \\
In \cite{Jacobi 45} Jacobi himself found his ``new multiplier''
for the  class of second-order ordinary differential
equations\footnote{This is not Jacobi's original notation.}
studied by Euler \cite{Euler} [Sect. I,  Ch. VI, \S\S 915 ff.]
(\ref{Jeq}). Indeed Jacobi derived that the multiplier of equation
(\ref{Jeq}) is given by:
\begin{equation}
M=e^{\varphi(t,x)}, \label{JM}
\end{equation}
as it is obvious from (\ref{Meq}). Consequently in our previous
paper \cite{nuctam_1lag}  we  derived a Lagrangian of the class of
equations (\ref{Jeq}) by means of (\ref{relMLo2}), i.e.
\begin {equation}
L=\frac{1}{2} e^{\varphi(t,x)}\dot x^2+f_3(t,x)+\frac{{\rm
d}}{{\rm d}t}G(t,x) \label{LagJ}
\end{equation}
with $f_3$ a function of $t$ and $x$ satisfying the following
equation:
\begin {equation}
  \frac{\partial f_3}{\partial x}
+e^{\varphi(t,x)} B(t,x)=0.
\end{equation}
Equation (\ref{careq}) is a particular example of the equation
considered by Jacobi. In fact from (\ref{JM}) and (\ref{LagJ}) we
derive:
\begin {equation}
 M=\frac{1}{\lambda x^2 + 1},
 \end{equation}
  and consequently
  \begin {equation}  L=\frac{\dot x^2}{2(\lambda x^2 + 1)}
 -\frac{a x^2}{2(\lambda x^2 + 1)}+\frac{{\rm
d}}{{\rm d}t}G(t,x).\label{Lcareq}
\end{equation}
This Lagrangian is known \cite{MatLak74}. Equation (\ref{careq})
does not possess any Lie point symmetries apart translation in
$t$. Therefore Noether's theorem applied to the autonomous
Lagrangian $L$ in (\ref{Lcareq}) yields the following first
integral:
\begin {equation}
I=\frac{a x^2 + \dot x^2}{2(\lambda x^2 + 1)}.
\end{equation}
Jacobi proved that in the case of a second-order differential
equation if one knows a first integral and a last multiplier then
the equation can be integrated by quadrature (a new Principle of
Mechanics, indeed) \cite{Jacobi 42}, \cite{Jacobi 44 a}.

Equation (\ref{NMeq}) is obtained by the symmetry reduction
transformation $x=u', t=u$ of the third-order equation:
\begin {equation}
u'''=-\frac{u'u''}{u},
\end{equation}
where $u(T)$ is a function of $T$. Equation (\ref{NMeq}) admits an
eight-dimensional Lie point symmetry algebra and therefore is
linearizable.
 In \cite{laggal} it was shown that if one knows several
(at least two) Lie symmetries of the second-order differential
equation (\ref{geno2}), i.e.
\begin {equation}
\Gamma_j =V_j(t,x)\partial_t+G_j(t,x)\partial_x, \quad j = 1,r,
 \label {gensym}
\end {equation}
 then many Jacobi Last Multipliers could be
derived by means of (\ref{2.8}), i.e.
\begin {equation}
{\displaystyle{\frac{1}{M_{nm}}}}=\Delta_{nm} = \mbox {\rm
det}\left [
\begin {array} {ccc}
1 &\dot x & F(t,x,\dot x)\\[0.2cm]
V_n &G_n &{\displaystyle{\frac{\d G_n}{\d t} -\dot x\frac{\d V_n}{\d t}}} \\[0.2cm]
V_m &G_m &{\displaystyle{\frac{\d G_m}{\d t} -\dot x\frac{\d V_m}{\d t}}}\\
\end {array}\right],\label {Mnm}
\end {equation}
with $(n,m=1,r)$, and therefore many Lagrangians  can be obtained
by means of (\ref{lagrint}). In particular, fourteen different
Lagrangians can be obtained if the equation admits an
eight-dimensional Lie point symmetry algebra. We do not look for
the fourteen Lagrangians of equation (\ref{NMeq}). Instead we use
equation (\ref{JM}) to find a Jacobi Last Multiplier and
consequently a Lagrangian. In fact from (\ref{JM}) and
(\ref{LagJ}) we derive:
\begin {equation}
 JLM=\frac{x^2}{t},
 \end{equation}
  and consequently
  \begin {equation}
   Lag=\frac{\dot x^2 x^2}{2 t}
 +\frac{{\rm
d}}{{\rm d}t}G(t,x).\label{LNMeq}
\end{equation}
If one applies Noether's theorem to $Lag$ then the following five
first integrals of equation (\ref{NMeq}) can be derived:
\begin{eqnarray}
FI_1&=& x^2 (x^2 - 2 x \dot x y + \dot x^2 t^2) \nonumber\\
 FI_2&=&\frac{x^2 \dot x ( - x + \dot x t)}
{2 t} \nonumber\\
FI_3&=&\frac{x^2 \dot x^2}{2 t^2}\label{NMI} \\
 FI_4&=& x (x - \dot x t)\nonumber\\
 FI_5&=&\frac{x \dot x}{t}\nonumber
\end{eqnarray}
We underline that the first integrals $FI_1$ and $FI_4$ could not
be derived if the gauge function $G(t,x)$ was assumed to be equal
to zero.

\section{Equations with space-dependent coefficients}

The class of equations (\ref{2})  has an obvious Jacobi Last
Multiplier and therefore Lagrangian if the following relationship
holds between $f(x)$ and $g(x)$:
\begin {equation}
\frac{{\rm d}}{{\rm
d}x}\left(\frac{g(x)}{f(x)}\right)=\alpha(1-\alpha)f(x) \label{3}
\end{equation} where $\alpha$ is any constant $\neq 1$. In fact if
(\ref{3}) holds then equation (\ref{2}) can be written as
\begin {equation}
\dot u+\alpha f(x) u=0 \label{ueq}
\end{equation} i.e.
\begin {equation}
f(x)=-\frac{1}{\alpha}\frac{{\rm d}}{{\rm d}t}(\log u),\quad\quad
{\rm with}\quad u=\dot x+\frac{g(x)}{\alpha f(x)},
\end{equation}
and thus a Jacobi Last Multiplier for equation (\ref{2}) is
\begin {equation}
M=\exp\left(\int f(x) dt\right) = \exp\left(-\frac{1}{\alpha} \int
d(\log u)\right)=u^{-1/\alpha} \label{Ms}\end{equation}
 and the corresponding Lagrangian is
\begin {equation}
L=u^{2-1/\alpha}+\frac{{\rm d}}{{\rm d}t}G(t,x)=\left(\dot
x+\frac{g(x)}{\alpha f(x)}\right)^{2-1/\alpha}+\frac{{\rm d}}{{\rm
d}t}G(t,x), \label{Ls}
\end{equation}
with $G(t,x)$ an arbitrary gauge function.\\
We note that this Lagrangian is autonomous and therefore it admits
at least the Noether point symmetry of translation in $t$, and
consequently the following first integral
\begin {equation}
In=\left(\dot x + \frac{g(x)}{\alpha f(x)}\right)^{1-1/
\alpha}\frac{\alpha f(x)\dot x - f(x)\dot x -
 g(x)}{\alpha^2 f(x)^2}\label{I2}
\end{equation}
We would like to remark that because of property (d) of the Jacobi
Last Multiplier, then we can obtain another Jacobi Last Multiplier
$\overline M=In M$ and consequently another Lagrangian of equation
(\ref{2}). We  will not pursue it here any further but one can
envision a deluge of Lagrangians obtained by  simply taking any
function of the first integral $In$ in (\ref{I2}) and multiplying
it for either $M$ in (\ref{Ms}) or $\overline M$, and so on ad
libitum.

\subsection{Examples}

It is very easy to obtain a Jacobi Last Multiplier and therefore a
Lagrangian for the  following Li\`enard type nonlinear oscillator:
 \begin {equation}
 \ddot x+kx\dot x+\frac{k^2}{9}x^3+\lambda x=0. \label{1}
 \end{equation}
In fact we know that $M=\exp(\int kx dt)$. Then if  one can put
the equation in the form
\begin {equation}
\dot u_1+\alpha k x u_1=0,\label{1u1eq}\end{equation}  with
$\alpha$ a constant to be determined, i.e.
\begin {equation}
kx=-\frac{1}{\alpha}\frac{{\rm d}}{{\rm d}t}(\log u_1)
\end{equation}
then
\begin {equation}
M=u_1^{-1/\alpha}. \end{equation} In the case of equation
(\ref{1}) we have
\begin {equation}
\dot u_1+\frac{1}{3}k x u_1=0, \quad \quad   {\rm with}\quad
u_1=\dot x +\frac{k}{3}x^2+\frac{3}{k}\lambda, \quad ({\rm
i.e.,}\quad \alpha=1/3).\end{equation}
 Therefore $M_1=u_1^{-3}$ and consequently\footnote{We  do not take
  into consideration any nonessential
multiplicative constant.}:
\begin {equation}
L_1=\frac{1}{u_1}+\frac{{\rm d}}{{\rm d}t}G(t,x)=\frac{1}{\dot x
+\frac{k}{3}x^2+\frac{3}{k}\lambda}+\frac{{\rm d}}{{\rm
d}t}G(t,x).
\end{equation}
Actually, we can derive another Lagrangian because substituting
$f(x)=kx, g(x)=\frac{k^2}{9}x^3+\lambda x $ into equation
(\ref{3}) yields two different $\alpha$, i.e.:
\begin {equation}
\frac{{\rm d}}{{\rm
d}t}\left(\frac{g}{f}\right)-\alpha(1-\alpha)f=0\quad
\Longrightarrow \quad 9\alpha^2 - 9\alpha + 2=0\quad
\Longrightarrow \quad \alpha_{1,2}=\frac{1}{3}, \frac{2}{3}
\end{equation}
The case $\alpha=\frac{1}{3}$ has been considered above. If we
substitute  $\alpha=\frac{2}{3}$ into equation (\ref{ueq}) then we
obtain $M_2=u^{-3/2}$ and consequently
\begin {equation}
L_2=\sqrt{u}+\frac{{\rm d}}{{\rm d}t}G(t,x)=\sqrt{\dot x
+\frac{k}{6}x^2+\frac{3}{2 k}\lambda}+\frac{{\rm d}}{{\rm
d}t}G(t,x).
\end{equation}
Equation (\ref{1}) admits an eight-dimensional Lie symmetry
algebra and therefore is linearizable. Moreover one can determine
at least twelve more Lagrangians \cite{CP07jlmmech}. We note that
$L_1$ admits one Noether point symmetry while $L_2$ admits three
Noether point symmetries.

A particular case of (\ref{1}) is the second-order Riccati
equation:
 \begin {equation}
 \ddot x+3x\dot x+x^3=0, \label{riceq}
 \end{equation}
 a member of the Riccati-chain \cite{Ames}. Equation (\ref{riceq}) is linearizable to a third-order linear equation
by the transformation $x=\dot V(t)/V(t)$, namely (\ref{riceq})
transforms into $\ddot V=0$. It also well-known that equation
(\ref{riceq}) is linearizable by means of a point transformation
because it admits an eight-dimensional Lie symmetry algebra
generated by the following operators:
\begin{eqnarray}
\Gamma_1&=&t^3(t x-2)\partial_t-t(x t-2) (x^2 t^2+2-2 x t)\partial_x\nonumber\\
\Gamma_2&=&x t^3\partial_t-(x t-1) (x^2 t^2+4-2 x t)\partial_x\nonumber\\
\Gamma_3&=&x t^2\partial_t-x (x^2 t^2+2-2 x t)\partial_x\nonumber\\
\Gamma_4&=&x t\partial_t-x^2 (x t-1)\partial_x\nonumber\\
\Gamma_5&=&x\partial_t-x^3\partial_x\\
\Gamma_6&=&\partial_t\nonumber\\
\Gamma_7&=&t\partial_t-x\partial_x\nonumber\\
\Gamma_8&=&t^2\partial_t-2(x t-1)\partial_x\nonumber
\end{eqnarray}
In order to find the linearizing transformation we have to look
for a two-dimensional abelian intransitive subalgebra \cite{Lie
12}, and, following Lie's classification of two-dimensional
algebras in the real plane \cite{Lie 12}, we have to transform it
into the canonical form
$$\partial_{\tilde x},\;\;\;\;\;\tilde t\partial_{\tilde x}$$ with
$\tilde t$ and $\tilde x$ the new independent
 and dependent variables, respectively. We found that one such subalgebra
  is that generated by $\Gamma_1$
and $\Gamma_9\equiv\Gamma_2-\Gamma_8$.
 Then, it is easy to derive that
 $$ \tilde t=  \frac{tx-1}{x(t x-2)},\;\;\;\;\;\tilde x=-\frac{x}{2t(t x-2)}$$
and equation (\ref{riceq}) becomes
\begin{equation}
{{\rm d}^2 \tilde x \over {\rm d}\tilde t^2}=0
\end{equation}
We can derive fourteen different Lagrangians by using (\ref{Mnm})
and (\ref{lagrint}). Two of these Lagrangians admit five Noether
symmetries, i.e.:
\begin {equation}
L_{56}=-\frac{1}{2(\dot x+x^2)}+\frac{{\rm d}}{{\rm d}t}G(t,x)
\label{rL56}
\end{equation}
and
\begin {equation}
L_{19}=-\frac{1}{2 t^4 (x^2 t^2+\dot x t^2-2 x t+2)}+\frac{{\rm
d}}{{\rm d}t}G(t,x).
\end{equation}
which are derived from
\begin {equation}
 JLM_{56}=-\frac{1}{(\dot x +x^2)^3},
\end{equation}
and
\begin {equation}
 JLM_{19}=-\frac{1}{(t^2 x^2+t^2\dot x-2t x+2)^3},
\end{equation}
 respectively.
 We remark that  $JLM_{56}$  can be also obtained from (\ref{ueq}). In
fact equation (\ref{riceq}) can be written as
\begin {equation}
\dot u+ 3 x \alpha  u=0, \quad \quad u=\dot x+x^2,\quad
\alpha=\frac{1}{3}.
\end{equation}
If one applies Noether's theorem to $L_{56}$ then the following
five first integrals of equation (\ref{riceq}) can be derived:
\begin{eqnarray}
I_1&=&\frac{(x^2  t^2 - 2  x  t + \dot  x  t^2 + 2)^2}{4  (x^2 +\dot  x)^2} \nonumber\\
I_2&=&\frac{(x^2  t^2 - 2  x  t + \dot  x  t^2 + 2)  (x^2  t - x +\dot  x  t)}
{2  (x^2 + \dot  x)^2} \nonumber\\
I_3&=&\frac{x^2 + 2  \dot  x}{2  (x^2 + \dot  x)^2}\\
 I_4&=&\frac{(x^2  t - x + \dot  x  t)^2}{(x^2 + \dot  x)^2}\nonumber\\
 I_5&=&\frac{x^2  t - x +\dot  x  t}{x^2 + \dot  x}\nonumber \label{ricI}
\end{eqnarray}
while Noether's theorem applied to $L_{19}$ yields:
\begin{eqnarray}
In_1&=&\frac{x^2  t - x + \dot  x  t}{x^2  t^2 - 2  x  t + \dot  x
t^2 +
2}\nonumber\\
In_2&=&\frac{(x^3  t - 2  x^2 - 2  \dot  x)  x  t + (\dot  x  t^2
+ 4) \dot  x
    + (2  \dot  x  t^2 + 3)  x^2}{(x^2  t^2 - 2  x  t + \dot  x  t^2 + 2)^2}\nonumber \\
In_3&=&\frac{(x^2  t - x + \dot  x  t)  (x^2 + \dot  x)}{(x^2  t^2
- 2 x  t
+ \dot  x  t^2 + 2)^2} \\
In_4&=&\frac{(x^2 + \dot  x)^2}{(x^2  t^2 - 2  x  t + \dot  x  t^2
+
2)^2}\nonumber \\
In_5&=&\frac{x^2 + 2  \dot  x}{2  (x^2  t^2 - 2  x  t + \dot  x
t^2 + 2)^2}\nonumber\label{ricIn}
\end{eqnarray}
We remark the importance of the gauge function $G(t,x)$. None of
the first integrals above, apart $I_3$ and $In_5$, could be
derived if the gauge function was assumed to be equal to zero.

\section{Final remarks}

The Lagrangian which admits the maximum number of Noether point
symmetries is that obtained by means of the Jacobi Last Multiplier
which comes from the determinant (\ref{Mnm}) with the two solution
symmetries, namely the two-dimensional abelian intransitive
subalgebra which yields the linearizing transformation. We may
infer that this is the physical Lagrangian. Also the Lagrangian
obtained by using the Jacobi Last Multiplier (\ref{Ms}) possesses
the maximum number of Noether point symmetries as shown in the
case of equation (\ref{3}). This may explain why Lagrangian
(\ref{Ls}) possesses  nice physical properties as shown in
\cite{Caretal05}.

In the present paper, we do not claim to have been exhaustive in
our presentation of the application of the Jacobi Last Multiplier
 for finding Lagrangians of any second-order differential
equation. Indeed we would like to encourage other authors to apply
Jacobi's method to their preferred equation.

\section*{Acknowledgements} This work was initiated while K.M.T.
was enjoying the hospitality of Professor M.C. Nucci and the
facilities of the Dipartimento di Matematica e Informatica,
Universit\`a di Perugia. K.M.T. gratefully acknowledges the
support of the Italian Istituto Nazionale Di Alta Matematica ``F.
Severi'' (INDAM), Gruppo Nazionale per la Fisica Matematica
(GNFM), Programma Professori Visitatori.


\begin{thebibliography}{99}
 \bibitem{Ames} Ames WF. Nonlinear ordinary differential equations
 in transport processes. New York: Academic Press; 1968.
\bibitem {Bianchi 18}
 Bianchi L. Lezioni sulla teoria dei gruppi continui finiti
di trasformazioni. Pisa: Enrico Spoerri; 1918.
\bibitem{Caretal05}
 Cari\~{n}ena JF.  Ra\~nada MF. and  Santander M.
 Lagrangian formalism for nonlinear second-order
Riccati systems: One-dimensional integrability and two-dimensional
superintegrability. J Math Phys, 2005;46:062703 (19pp).
\bibitem{Euler}
 Euler L. E366 -  Institutionum calculi integralis, Volumen
Secundum. Petropli impensis academiae imperialis scientiarum;
1769.
\bibitem{NorMar}
Euler N and Euler M. Sundman symmetries of nonlinear second-order
and third-order ordinary differential dquations. J Nonlinear Math
Phys, 2004;11:399-421.
\bibitem{Jacobi 42}
 Jacobi CGJ. Sur un noveau principe de la m\'ecanique
 analytique.  C R Acad Sci Paris, 1842;15:202--205.
 \bibitem {Jacobi 44 a}
Jacobi CGJ. Sul principio dell'ultimo moltiplicatore, e suo uso
come nuovo principio generale di meccanica.  Giornale Arcadico di
Scienze, Lettere ed Arti, 1844;99:129--146.
\bibitem {Jacobi 44 b}
Jacobi CGJ.  Theoria novi multiplicatoris systemati \ae quationum
differentialium vulgarium applicandi. J Reine Angew Math, 1844;
27:199--268.
\bibitem {Jacobi 45}
Jacobi CGJ. Theoria novi multiplicatoris systemati \ae quationum
differentialium vulgarium applicandi. J Reine Angew Math,
1845;29:213-279 and 333-376.
 \bibitem{JacobiVD}  Jacobi CGJ.
 Vorlesungen \"uber Dynamik. Nebst f\"unf hinterlassenen Abhandlungen
 desselben herausgegeben  von A.Clebsch. Berlin: Druck und Verlag von Georg
 Reimer; 1886.
 \bibitem{Lie1874}  Lie S. Veralgemeinerung und neue Verwerthung der Jacobischen
Multiplicator-Theorie. Fordhandlinger i Videnokabs - Selshabet i
Christiania, 1874; pp. 255-274.
 \bibitem {Lie 12}
 Lie S. Vorlesungen \"uber Differentialgleichungen mit
bekannten infinitesimalen transformationen. Leipzig: Teubner;
1912.
\bibitem{MatLak74}
Mathews P M and Lakshmanan M. On a unique non-linear oscillator. Q
Appl Math, 1974;32:215-218
  \bibitem{musetal08}
 Musielak ZE, Roy D and Swift LD.  Method to derive Lagrangian and
Hamiltonian for a nonlinear dynamical system with variable
coefficients. Chaos, Solitons \& Fractals, 2008;38:894-902.
\bibitem{musetal08_2}
Musielak ZE. Standard and non-standard Lagrangians for dissipative
dynamical systems with variable coefficients. J. Phys A: Math.
Theor, 2008;41:055205 (17pp)
\bibitem{Noether}
  Noether E. Invariante Variationsprobleme. Nachr d K\"onig Gesellsch d Wiss zu
 G\"ottingen Math-phys Klasse, 1918;235-257.
\bibitem{man2}
 Nucci MC. Interactive REDUCE programs for calculating Lie
point, non-classical, Lie-B\"{a}cklund, and approximate symmetries
of differential equations: manual and floppy disk, in CRC Handbook
of Lie Group Analysis of Differential Equations, Vol. 3: New
Trends in Theoretical Developments and Computational Methods, ed.
 Ibragimov NH. Boca Raton: CRC Press; 1996, pp. 415-481.
 \bibitem{ennity}
Nucci MC and Leach PGL.
 Jacobi's last multiplier and the complete symmetry group of the Euler-Poinsot
 system.  J Nonlinear Math Phys,  2002;9S2:110-121.
 \bibitem{JLM}
 Nucci MC and Leach PGL.  Jacobi's last multiplier and symmetries
 for the Kepler problem plus a lineal story. J Phys A: Math Gen,
 2004;37:7743-7753.
\bibitem{jlm05}
Nucci MC.  Jacobi last multiplier and Lie symmetries:
 a novel application of an old relationship.
  J Nonlinear Math Phys, 2005;12:284-304.
   \bibitem{jlmcsepeq}
 Nucci MC and Leach PGL. Jacobi's last multiplier and the complete
 symmetry group of the Ermakov-Pinney equation. J Nonlinear Math
 Phys, 2005;12:305-320.
 \bibitem{gallipoli04}
 Nucci MC.  Let's Lie: a miraculous haul of fishes. Theor Math
 Phys, 2005;144:1214-1222.
\bibitem{PVI}
Nucci MC and Leach PGL.  Fuch's solution of Painlev\'e VI equation
by means of Jacobi last multiplier. J Math Phys, 2007;48:013514
(7pp)
  \bibitem{gallipoli06}
 Nucci MC.  Jacobi last multiplier, Lie symmetries, and hidden
 linearity: ``goldfishes" galore. Theor Math Phys,
 2007;151:851-862.
\bibitem{laggal}
Nucci MC and Leach PGL.  Lagrangians galore. J Math Phys,
2007;48:123510 (16pp).
 \bibitem{PInceeq}
  Nucci MC. Lie symmetries of a Panlev\'e-type equation without Lie
  symmetries. J Nonlinear Math Phys, 2008;15:205-211.
  \bibitem{jlmschqm}
 Nucci MC and Leach PGL. Gauge variant symmetries for the
 Schr\"odinger equation.  Il Nuovo Cimento B, 2008;123:93-101.
\bibitem{CP07Rao1JMP}
Nucci MC and Leach PGL. Jacobi last multiplier and Lagrangians for
multidimensional linear systems. J Math Phys, 2008;49:073517
(8pp).
\bibitem{CP07jlmmech}
Nucci MC and Leach PGL. The Jacobi last multiplier and
applications in Mechanics.  Phys  Scr, 2008;78: (to appear)
\bibitem{CP07Naill}
Nucci MC and Leach PGL. An old method of Jacobi to find
Lagrangians. ArXiv:0807.2796v1 [nlin.SI] (2008)
\bibitem{nuctam_1lag}
 Nucci MC and   Tamizhmani KM.  Using an old method of Jacobi
to derive Lagrangians: a nonlinear dynamical system with variable
coefficients. ArXiv:0807.2791v1 [nlin.SI] (2008)
\bibitem{Whittaker}
 Whittaker ET.  A Treatise on the Analytical Dynamics of
Particles and Rigid Bodies. Cambridge: Cambridge University Press:
1988 (First ed. 1908).


\end{thebibliography}
 \end{document}